\documentclass[a4paper,12pt]{article}

\usepackage{graphicx}
\usepackage{color}

\title{`Flow \& Jam' of frictional athermal systems under shear stress}

\author{Raffaele Pastore$^{\rm a,b}$ $^{\ast}$\thanks{$^\ast$Corresponding 
author. Email: pastore@na.infn.it \vspace{6pt}}, 
Massimo Pica Ciamarra$^{\rm a,b}$ and
Antonio Coniglio$^{\rm a,b}$
\\\vspace{6pt} $^{\rm a}${\em{Dip. Scienze Fisiche, Universit\`a di
Napoli Federico II, Italy}}; \\
$^{\rm b}${\em{CNR--SPIN}}
\\\vspace{6pt}
}

\date{ September 2011}

\begin{document}

\maketitle

\begin{abstract}
We report recent results of molecular dynamics simulations of frictional
athermal particles at constant volume fraction and
constant applied shear stress, focusing on a range of control parameters where
the system first flows, but then
jams after a time $t_{jam}$. On decreasing the volume fraction, the mean jamming
time diverges, while its
sample fluctuations become so large that the jamming time probability
distribution $P(t_{jam})$ becomes a 
power-law. We obtain an insight on the origin of this phenomenology focusing on
the flowing regime, which is characterized by the
presence of a clear correlation between the shear velocity and the mean number
of contacts per particles $Z$, whereby small velocities
occur when $Z$ acquires higher values. 
\end{abstract}

\section{Introduction}
The non-equilibrium transition from a fluid--like state to a disordered
solid--like state, known as jamming transition, 
occurs in a wide variety of physical systems, such as colloidal suspensions and
molecular fluids. 
Its widespread occurrence suggested to introduce a `jamming phase diagram', with
axes temperature, density and shear
stress~\cite{nagel,Trappe,o'hern,Coniglio}, able to describe the jamming
transition of all of these systems.
In this framework, jamming occurs at small temperature, high density, and small
external shear stress.
Frictionless athermal systems~\cite{nagel} can be described by the zero
temperature plane of the jamming diagram: a
transition line is expected to separate a jammed phase, which consists in
non--ordinary solid~\cite{Cates1, Cates2}, 
and a flowing phase, where only steady states are
observed~\cite{PCC09,Ohern06}. 
Such a simple picture becomes more involved in the presence of friction, which
must be taken into account in order to
properly describe granular (macroscopic) materials 
~\cite{Zhang, Cates3, Makse, van Hecke, Magnanimo,
Behringer,Grebenkow,arxiv}.

We have recently investigated the effect of friction on the jamming
phenomenology performing Molecular Dynamics (MD) simulations 
of a frictional granular system at constant volume fraction and constant applied
shear stress~\cite{Grebenkow,arxiv}. 
In this set--up the jamming phenomenology is very rich, due to the occurrence of
four possible {\it phases}, `Flow', `Flow \& Jam', `Slip \& Jam' and 'Jam' (as
found on increasing the
density or decreasing the applied shear stress). In the `Flow \& Jam' and in the
`Slip \& Jam' phase, the
system reaches an apparently steady flowing state before jamming, or jams after
a tiny displacement, respectively.
In this paper, we present recent results regarding the `Flow \& Jam' regime,
where the system jams after flowing for a time $t_{jam}$. 
We describe both dynamical properties of the system, such as the jamming time,
as well as structural ones, such as the evolution of mean
contact number. Connections and correlations between static and dynamics
quantities suggest which is the physical mechanisms responsible of
the `Flow \& Jam' phenomenology.

\section{The investigated system}
We perform MD simulations of the system considered in~\cite{Grebenkow,arxiv}. 
Monodisperse spherical grains of mass $M$ and diameter $D$ are enclosed in a box
of dimension $l_x = l_y = 16D$, and $l_{z} = 8\textit{D}$. Periodic boundary
conditions are used along $x$ and $y$, while the size of the vertical dimension
is fixed and chosen to be comparable to that of recent experiments~\cite{Pine,
Daniels, Behringer2}. The upper and lower boundary surfaces of the box consist
in rough ``virtual'' plates: each plate is made by a collection of particles
that move as a rigid object. The bottom plate has an infinite mass, and is therefore fix,
while the top one has a mass equal to the sum of the masses of its particles (roughly $l_xl_y$).
The top plate is subject to a shear stress $\sigma_{xz}$ ($\sigma_{xz} = \sigma$ from now on).

Grains interact via the standar linear spring-dashpot model.  Two particles $i$ and
$j$, in positions ${\bf r}_{i}$ and ${\bf r}_{j}$, with linear velocities ${\bf v}_{i}$ and ${\bf v}_{j}$,
and angular velocities $\omega_{i}$ and $\omega_{j}$ , interact if in contact,
i.e., if the quantity 
$\delta_{ij}=D-|{\bf r}_{ij}|$ is positive. $\delta_{ij}$ is called the
penetration lenght, and ${\bf r}_{ij}={\bf r}_{i}-{\bf r}_{j}$ is the distance
between the center of the particles $i$ and $j$. The interaction force has a
normal component ${\bf F}_{n_{ij}}$  and a tangential one ${\bf F}_{t_{ij}}$,
the both having an elastic and a dissipative component:
\begin{equation}
 {\bf F}_{n_{ij}} = -k_{n} \delta_{ij} {\bf n}_{ij} - \gamma_n m_{eff} {\bf
v}_{{\bf n}_{ij}}
\end{equation}

\begin{equation}
 {\bf F}_{t_{ij}} = -k_{t} {\bf u}_{t_{ij}} - \gamma_t m_{eff} {\bf v}_{{
t}_{ij}}
\end{equation}

where $k_n$ and $k_{t}$ are elastic moduli, while $\gamma_n$ and $\gamma_t$
account for dissipative caharacter of the normal and tangential component
rispectively. ${\bf n}_{ij} = {\bf r}_{ij}/|{\bf r}_{ij}|$, ${\bf v}_{n_{ij}} =
[({\bf v}_{i} - {\bf v}_{j}) \cdot {\bf n_{ij}}] {\bf n}_{ij}$, ${\bf v}_{{\bf
t}_{ij}} = {\bf v}_{{ij}} - {\bf v}_{{\bf n}_{ij}}$ and $m_{eff}$ is the
effective mass. 
${\bf u}_{\bf t_{ij}}$, set to zero at the beginning of a contact, measures the
shear displacement during the lifetime of a contact. Its time evolution is fixed
by ${\bf v}_{{\bf t}_{ij}}$  and $\omega_{i}$ and $\omega_{j}$, as
described in Ref.~\cite{Silbert}. The presence of tangential forces implies the
presence of torques, $\tau_{ij} = -1/2 {\bf r}_{ij}\times {\bf F}_{ij}$. The
shear displacement is set to zero both when a contact finish
($\delta_{ij}<0$), where $\mu$ is the coefficient of static friction.
We use the value of the parameters of~\cite{Silbert}: $k_n=2~10^{5}$,
$k_{t}/k_{n}=2/7$, $\gamma_{n}=50$, $\gamma_{t}/\gamma_{n}=0$. Lenght, masses
and times are expressed in units of $D$, $m$ and $\sqrt{m/k_n}$.

The volume fraction $\varphi$ represents the volume occupied by the grains
divided by the volume of the container, i.e. $\varphi=Nv_{0}/V_{0} +
\Delta\phi$, where $V_{0}=l_{x}l_{y}l_{z}$ is the volume of the system and
$v_{0}=1/6\pi D^{3}$ is the volume occupied by a single grain and
$\Delta\phi=0.021$ is a corrective a term which takes into account the effect of
the rough plates protruding into the system. By changing the number of particles
we vary the volume fraction.

The initial state is prepared setting to zero the friction coefficient
~\cite{Song}, randomly placing the particles and then inflating them until the
desired volume fraction is obtained; such a protocol is a short-cut of
experimental procedures with which it is possible to generate very dense
disordered states of frictional systems, such
as oscillations of high frequency and small amplitude~\cite{Gao}. After
preparing the system we `switch on' friction, and follow the time evolution of the system under the action of 
the constant shear stress .
We  investigate the
evolution of the system
focusing on a value of
Coulomb friction coefficient equal to~$\mu=0.1$ 
while $\varphi$ and $\sigma$ are the variable control parameter. All the data
shown here refers to points ($\varphi , \sigma$) falling in the `Flow \& Jam'
region of the Jamming--phase--diagram~\cite{arxiv}. For each point we perform
$50$--$100$ different runs, starting from different initial conditions. 
In the `Flow \& Jam' region, the steady state last a long time at small volume fractions, where
the time $t_{jam}$ the system flows before jamming is large. This is the region where
the `Flow \& Jam' is more surprising, we have investigated in detail performing simulations lasting up to a time $T=5~10^{4}$.
At higher volume fraction $t_{jam}$ is small, and the identification of the steady state becomes more difficult.

\section{Flow and Jam region}
\begin{figure}[t!]
\begin{center}
\includegraphics*[scale=0.3]{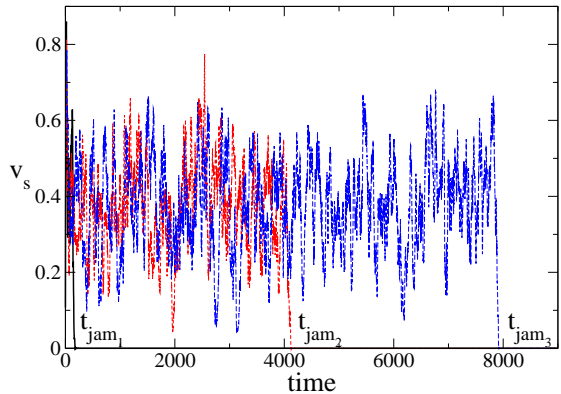}
\end{center}
\caption{\label{fig:velocity}
(Color online) Velocity of the upper plate $v_{s}$ as a function of time for
three simulations at $\phi \simeq 0.627$ and $\sigma
=2~10^{-3}$.  We marked the values of the jamming times, $t_{jam}$, which differ
by more than three decades.
}
\end{figure}

\begin{figure}[t!]
\begin{center}
\includegraphics*[scale=0.3]{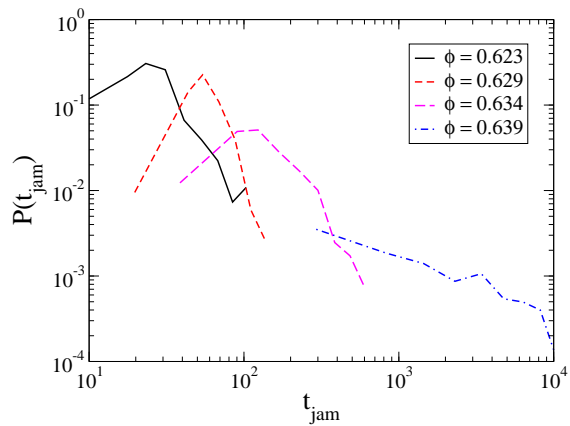}
\end{center}
\caption{\label{fig:probability}
(Color online) Probability distribution $P(t_{jam})$ of the jamming time for
$\sigma =2~10^{-3}$ and different values of the volume fraction,
as indicated.
}
\end{figure}

\subsection{Jamming times}
In the `Flow \& Jam' phase of sheared granular systems~\cite{arxiv}, the system
first flows with a constant velocity as in a steady flowing
phase, but then suddenly jams after a time $t_{jam}$. Such a phenomenology is
shown in Fig.~\ref{fig:velocity}, where we report the
velocity $v_{s}(t)$ of the upper plate as a function of time for three
simulations at $\phi=0.627$ and $\sigma=2~10^{-3}$. 
The average jamming time $\langle t_{jam}\rangle$ depends on the volume
fraction. It diverges as a power law on
decreasing the volume fraction, at a critical volume fraction
$\phi_{j1}(\sigma,\phi)$, and vanishes on increasing the volume fraction, 
at a critical value $\phi_{j2}(\sigma,\phi)$.
The lines $\phi_{j1}(\sigma,\phi)$ and $\phi_{j2}(\sigma,\phi)$ define the
boundary of the
`Flow \& Jam' phase in the volume fraction, shear stress and friction jamming
phase diagram for frictional particles~\cite{arxiv} .

However, the jamming time is subject to large fluctuations. For instance, the
three simulations shown in Fig.~\ref{fig:velocity} jam at 
very different times, even though they differ only in the initial
configuration. 
Such an observation suggested to study the sample fluctuations of the jamming
time, which we show in
Fig.~\ref{fig:probability} for $\sigma = 2~10^{-3}$ and three different values
of the volume fraction. 
While at high volume fractions $P(t_{jam})$ is peaked, meaning that the jamming
time is well defined, on decreasing the volume fraction
the distribution moves to larger times, and at the same times changes shape,
becoming well described by a power law.

In order to understand the origin of the `Flow \& Jam' phenomenology and of the
large fluctuations of the jamming time, which are 
not described by current rheological models such as those based on the inertial
number~\cite{Pouliquen} or on rate and state
equations~\cite{RS1,RS2}, we have analyzed the evolution of the micro-structure
of the system in the flowing regime and the micro-mechanics
of the jamming process, as described below.

\begin{figure}[t!]
\begin{center}
\includegraphics*[scale=0.3]{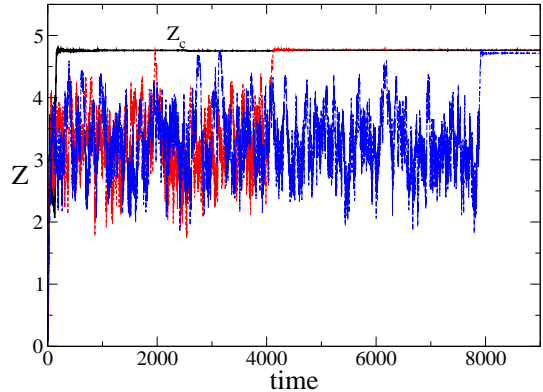}
\end{center}
\caption{\label{fig:Zt}
(Color online) Mean number of contacts per grain $Z$ as a function of time, for
the simulations shown in Fig.\ref{fig:velocity}. 
The system jams when $Z$ reaches a critical value $Z_{c}$.
}
\end{figure}

\begin{figure}[t!]
\begin{center}
\includegraphics*[scale=0.3]{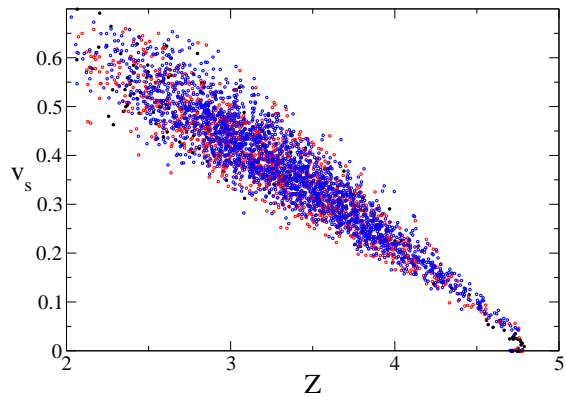}
\end{center}
\caption{\label{fig:VZ}
(Color online) Parametric plot of the velocity of the top plate versus the mean
number of contacts, obtained from the data shown in
Fig.s~\ref{fig:velocity} and \ref{fig:Zt}. The collapse indicates the existence
of a correlation between $Z$ and $v_s$: the higher $Z$,
the slower the system. 
}
\end{figure}

\subsection{Fluctuations of the micro-structure of the system}
Fig.~\ref{fig:Zt} illustrates the time evolution of the mean number of contacts
per grain, $Z(t)$, for the same simulations considered
in Fig.~\ref{fig:velocity}.
At the beginning of the simulations, as a consequence of the considered
preparation protocol, $Z(t = 0) = 0$.
$Z(t)$ rapidly increases as the system start flowing, and then fluctuates around
a constant value in the following steady flowing phase.
A comparison between Fig.~\ref{fig:velocity} and Fig.~\ref{fig:Zt} suggests the
presence of a correlation between the shear
velocity $v_s(t)$ and the mean contact number $Z(t)$, whereby large values of
$Z$ occurs when the shear velocity is small, and conversely.

Fig.~\ref{fig:VZ}
shows
a parametric plot of $v_{s}(t)$ versus $Z(t)$. 
In such a representation, the data 
from simulations  
characterized by very different values of
the jamming time 
display a correlate behaviour. 

\begin{figure}[t!]
\begin{center}
\includegraphics*[scale=0.3]{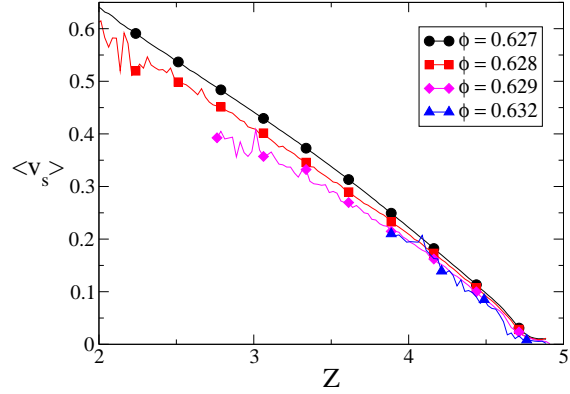}
\end{center}
\caption{\label{fig:Vmed_Z}
(Color online)  
Dependence of the average shear velocity $\langle v_{s} \rangle $ as a function
of the mean number of contacts per grain  at $\sigma =2~10^{-3}$, for the
indicated values of the volume fraction.
}
\end{figure}

\begin{figure}[t!]
\begin{center}
\includegraphics*[scale=0.5]{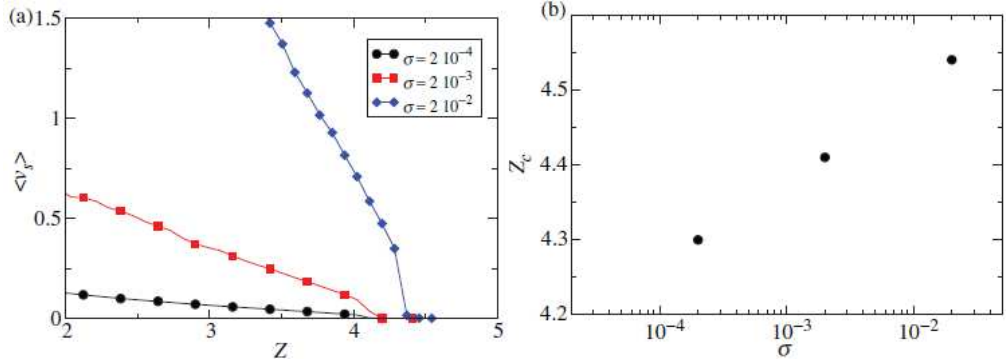}
\end{center}
\caption{\label{fig:Vmed_Z_svar}
(Color online) At $\phi = 0.629$, \textbf{a)} average shear velocity $\langle
v_{s} \rangle $ versus mean number of contacts per grain for
the indicated values of the shear stress, and \textbf{b)} jamming critical mean
contact number $Z_{c}$ as a function of the applied shear stress.
}
\end{figure}

Note that the fluctuations of the shear velocity decreases as the mean number
of contacts increases, suggesting the presence of a well defined critical mean
number of contacts $Z_c$ at which jamming occurs, in agreement with  the results
of Fig.~\ref{fig:Zt}.
We plot in Fig.~\ref{fig:Vmed_Z} the averaged shear velocity as a function of
the mean number of contacts, $\langle v_{s}(Z) \rangle$,The normal component is
given by
for the indicated values of the volume fraction. The figure suggests that $Z_c$
is almost independent on
the volume fraction, consistently with the existence of a volume fraction range
where granular systems with equal mechanical properties can
be prepared~\cite{arxiv}. On the contrary, as shown in
Fig.s~\ref{fig:Vmed_Z_svar}a,b $Z_c$ increases with the applied shear stress.  
Considering that the shear modulus is expected to increase with the mean contact
number~\cite{o'hern}, this result suggests that the shear modulus of a jammed
system does not simply depend on its volume
fraction, but also on the applied stresses which caused jamming, or more
generally on the preparation procedure.

\section{Jamming mechanism}
Here we propose a qualitative mechanism to explain the origin of the `Flow \&
Jam' phenomenology, based on the
behavior of the shear velocity $\langle v_{s}(t) \rangle $ and of the mean
contact number $Z(t)$, as well as on the
dependence of $t_{jam}$ and $Z_{c}$ on the control parameters. The starting
point is the well known dilatancy phenomenon observed in
granular systems~\cite{Reynolds1, Reynolds2, Kabla}, which is the tendency of
flowing particulate systems to dilate. 
At constant pressure and constant shear strain rate a dilation is actually
observed, the larger the greater the shear velocity~\cite{Blair}. 
At constant volume, which is the case considered here, dilation is obviously
forbidden. 
This leads to a impeded dilatancy which may explain the observed phenomenology. 

While flowing, the system visits different microscopic configurations, each one
having a typical mean number of contact $Z$. When $Z$ is
small, particles exert a small resistance to the applied stress, the shear rate
increases and the system tries to dilate. This leads to a
configuration with a larger mean number of contacts, exerting a larger
resistance, which causes the system to decelerate. 
The existence of such a feedback mechanism is suggested by the correlations
between $Z$ and $v_s$ illustrated in Fig.~\ref{fig:VZ}.
The impeded dilatancy appears therefore responsible for the fluctuations of the
mean number of contacts. 
The flowing system jams as a result of a large fluctuation of the mean number of
contacts $Z$, which reaches the critical value $Z_c$ 
corresponding to configurations able to sustain the applied stress.
How frequent are these fluctuations? We expect these fluctuations to be more
rare when the volume fraction is small, simply because there are
fewer configurations able to sustain the applied stress (i.e. with $Z = Z_c$):
this explains why the jamming time
increases as the volume fraction decreases. Also, one expects that when the
volume fraction is smaller than a threshold value, there are no 
configurations with $Z = Z_c$, which explains why the jamming time diverges
decreasing the volume fraction. 


\section{Conclusions}
In this manuscript, we focused on a region of the control parameters where
frictional granular systems jam after flowing
with a constant velocity, and described a
possible mechanism able to explain the observed behavior, The mechanism is based
on the notion of impeded
dilatancy and on the presence of correlations between structure and dynamics of
the system in the flowing regime.
An open question ahead is the explanation of the power-law like distribution of
the jamming times at small volume fractions, an
indication of the presence of correlations in the dynamics of sheared systems,
which needs to be clarified.


\section{Aknowledgments}
We acknowledge computer resources from the University of Naples Scope grid
project, CINECA, CASPUR and DEISA.

\end{document}